\documentstyle[pra,aps,epsf,preprint,rotating]{revtex}
\tightenlines
\begin{document}

\title{Two Large-Area Anode-Pad MICROMEGAS Chambers as the basic
elements of a Pre-Shower Detector}

\author{L.~Aphecetche, H.~Delagrange, D.~G. d'Enterria, M.~Le Guay,
X.~Li\thanks{Present address: China Institute of Atomic Energy,
P.O. Box 275, 102413 Beijing, China.}, \\
G.~Mart\'{\i}nez\thanks{Corresponding author: martinez@in2p3.fr },
M.J.~Mora, P.~Pichot, D.~Roy and Y.~Schutz }

\address{SUBATECH \\
(Ecole des Mines de Nantes, IN2P3/CNRS, Universit\'e de Nantes ), \\
 BP 20722, 44307 Nantes Cedex 3, France}

\maketitle

\begin{center}{Abstract}\end{center}

The design of a detector based on MICROMEGAS (MICRO MEsh GAseous Structure)
technology is presented.
Our detector is characterized by a large active area of 398\( \times \)281 mm\( ^{2} \),
a pad read-out with 20\( \times \)22 mm\( ^{2} \) segmentation,
and an uniform amplification gap obtained by insulating spacers (100 \( \mu  \)m high and 200 \( \mu  \)m in diameter).
The performances of several prototypes have been evaluated under irradiation with secondary
beams of 2 GeV/c momentum charged pions and electrons.
We consider such a detector as the basic element for a pre-shower
detector to equip the PHOton Spectrometer (PHOS) of the ALICE experiment.
Its assets are modularity, small amount of material,
robustness and low cost.

\section{Introduction}

The main goal of experiments at relativistic heavy-ion colliders,
like RHIC at Brookhaven and LHC at CERN, is to produce and study,
in the laboratory, the primordial matter of the universe which is predicted to
consist of a plasma of deconfined quarks and gluons (QGP). Among
the various considered probes and related observables (strange
hadrons, heavy mesons, di-electrons, etc.), direct photons will
explore the partonic phase during the equilibration process, as
well as the QGP phase itself \cite{Cheu94,Albr96,QM99}. In the ALICE (A
Large Ion Collider Experiment) experiment at CERN \cite{ALICE95},
direct photons will be detected and identified with PHOS (PHOton
Spectrometer) \cite{PHOS99}. PHOS is a highly segmented
electromagnetic calorimeter (EMCA) which will consist of 17280
lead-tungstate (PbWO\( _{4} \)) crystals, associated to a charged
particle veto detector (CPV). Each crystal (2.2\( \times
\)2.2\(\times \)18 cm\(^{3} \), 20 radiation lengths) is coupled to a
PIN-diode associated to a low noise pre-amplifier. The PHOS
spectrometer will be positioned at the bottom of the ALICE magnet,
4.6 meters away from the interaction point. It will cover the
pseudo-rapidity range from -0.12 to 0.12 and the azimuthal-angle
domain from 45\( ^{\circ } \) to 135\( ^{\circ } \). Rejection of
charged hadronic showers will be performed with the CPV detector,
positioned in front of the PHOS crystals.

Recent analyses of the direct photon production in heavy-ion
collisions at SPS energies \cite{Albr96,QM99,Agga99} have pointed
out the limits of the photon identification power resulting from two kinds of
contaminations: i) the showers induced by neutral
hadrons (mainly \( n \) and \( \bar{n} \)), and ii) the overlap of
showers in high multiplicity events. These contaminations are
expected to become much more severe at LHC energies. In order to
properly minimize these effects, we have designed a
pre-shower detector to be associated with PHOS. This detector
noticeably improves the photon identification power and allows for
a direct measurement of the electromagnetic shower vertex. The PPSD
(PHOS Pre-Shower Detector, Fig. 1) consists
of two distinct MICROMEGAS (MICRO MEsh GAseous
Structure) gas counters: the Charged Particle Veto (CPV) and the
Photon Conversion (PC) chambers, sandwiching a passive Pb converter.

Among the new micro-pattern detectors exploiting avalanche multiplication in narrow gaps \cite{Saul98},
MICROMEGAS technology \cite{Giom96} appears as a very promising option in terms of
performances, robustness, reduced amount of material, and cost (see
e.g. \cite{Giom98,Char98,Barc99,Cuss98,Micr99,Micr00}).
For our specific needs, we justify the choice of the MICROMEGAS technology
with the following considerations:

\begin{enumerate}
\item It provides the modularity well suited to cover areas
of a few squared meters.
\item It is mechanically robust and resistant to electrical discharges.
\item The small lateral development of the electron cloud in the
amplification gap, prevents from significant overlaps in high multiplicity events.
\item The small amount of material minimizes photon-conversion and
nuclear interactions within the detector.
\end{enumerate}

Using our PPSD design, the identification capabilities of PHOS will be
improved as follows:

\begin{enumerate}
\item Muons will be identified as minimum ionizing particles (MIPs) in the CPV, PC
and EMCA detectors.
\item About half of the charged hadrons (\( \pi ^{+} \), \( \pi ^{-} \), K\( ^{+} \),
K\( ^{-} \), etc.) will be detected as MIPs in the CPV, the PC, and the calorimeter.
The other half of charged hadrons will develop a
hadronic shower in the calorimeter.
\item Photons will be unambiguously identified by three features: the absence of a hit in the CPV,
a hit in the PC detector, when the photon is converted into an $e^+-e^-$ pair within the absorber,
and an electromagnetic shower in the
calorimeter. The identification efficiency will depend on the
thickness of the passive converter. The optimal thickness results
from a compromise between the photon identification efficiency of
the PPSD and the photon energy resolution of the spectrometer. From
simulations \cite{Aliroot99}, we deduced that the best trade-off is
achieved for a thickness equivalent to one radiation length.
Additionally, from the hit position in the PC detector, the shower
vertex will be directly measured, providing information to resolve
the overlapping showers expected in high multiplicity events.
\item Neutral hadrons will trigger most of the time nor the CPV, neither the PC detector
(the intermediate Pb passive converter represents only 5\% of the
nuclear interaction length).
\item Electrons and positrons will be identified by a hit in the CPV
and PC and by an electromagnetic shower developing in the EMCA.
Additional improvement of the electron and positron identification
power could be achieved by considering the deposited energy in the
PC detector.
\end{enumerate}

In the present report, the performances of large-area anode-pad
MICROMEGAS detectors, irradiated with a 2 GeV/c pion and electron
beam, are presented in terms of pad response to MIP, gain in the
amplification gap and detector efficiency. In addition, we have
studied the electrical discharge probability per ionizing particle
as a function of the detector gain, in order to probe the
reliability of this technology in large multiplicity experiments.
Finally, we have studied the electron and pion response functions
of a pre-shower detector prototype.

\section{Description of a Large-Area Anode-Pad MICROMEGAS Chambers}

The MICROMEGAS technology \cite{Giom96,Giom98} consists of an
asymmetric two-stage parallel-plate gas-detector. The first stage,
a 3~mm thick conversion gap, is separated from the 100~\( \mu \)m
thick amplification gap, by a micro-mesh foil resting on
insulating spacers. The amplification gap is closed at the bottom
by the pad anode surface. Such a configuration allows to establish
simultaneously a very high electric field in the amplification
region (\( \sim \) 50~kV/cm) together with a low electric field in
the drift region (\( \sim \) 2 kV/cm), by applying suitable
voltages between the three electrodes (cathode-mesh-anode). When a
charged particle traverses the conversion gap, it generates primary
electrons which are subsequently multiplied in the small
amplification gap. The associated ion cloud is quickly collected on
the micro-mesh layer generating a relatively fast signal, whereas
only a small part of the ion cloud penetrates into the conversion
region. The amplified electron cloud is collected on the anode
providing a fast electric signal.

Based on this principle, we have developed several large-area
prototypes with pad read-out (Fig. 2 and Fig. 3). The
characteristics of the chambers are the following:

\begin{enumerate}
\item {\bf Anode}: The total area of the anode plane is \( 480\times 300 \)
~mm\( ^{2} \), providing an active area of \( 398\times 281 \)~mm\(
^{2} \). The anode electrode consists of a 1.0 mm thick printed circuit
board. Its inner surface is segmented in rectangular 20 \(
\times \) 22 mm\(^{2} \) gilded copper pads, and
the signal is collected on the other side through strips to the
connectors. The inter-pad width is 100 \( \mu \)m and the total
number of pads is 256. Each pad is pierced by a conductive pine
hole of 300\( \, \mu \)m in diameter to allow for readout through the
board. The rigidity of the board is obtained by sticking an additional 3 mm
thick board made of low \( X_{0} \) composite material (EPOXY glass
and ROHACELL). To keep the small amplification gap as uniform as
possible, well-calibrated micro-spacers (100 \( \mu \)m high and
200 \( \mu \)m in diameter) are deposited on the pads with a pitch
of 2 mm in both directions.
\item  {\bf Cathode}: The cathode consists of a 9 \( \mu  \)m
gilded copper layer, glued on a 3 mm thick plate made of composite
material and sandwiched between two 300 \( \mu \)m thick GI180
boards. The top cathode plane is glued to a 6 mm thick Plexiglas
frame of \( 410\times 294 \)~mm\( ^{2} \).
\item  {\bf Mesh}: The original electro-formed micro-mesh\footnote{
BMC Industries, 278 East 7th Street, St. Paul, MN 55101, USA.}
consists of a 3 \( \mu \)m thick grid of 22''\( \times \)22'' made of pure Ni. The 39~\( \mu \)m squared holes grid are
outlined by a 11~\( \mu \)m thick border of Ni in steps of 50~\(\mu
\)m, i.e. 500 LPI (``Lines Per Inch''). The optical transparency
reaches 59\%. The micro-mesh is stretched on the Plexiglas frame
which defines the 3 mm thick conversion gap between the micro-mesh
and the cathode plane.
\item The micro-mesh and cathode assembly is placed on top of the
micro-spacers of the anode plane. A 2 mm thick composite-material
top-lid covers the whole system to ensure the gas tightness of the
chamber.
\item The gas mixture flows through the detector at a pressure slightly
above atmospheric pressure.
\item The signals of the individual pads are collected, through the
metallic hole, at the backplane of the anode plane by individual
strips and transported to the front-end electronics located on the
two opposite sides of the detector.
\item The total thickness of the detector throughout the active
detection area, is 13 mm corresponding to 1.76\% of \( X_{0} \).
\end{enumerate}

\section{Experimental setup for the beam tests}

Our prototype of the PPSD detector was tested in the
T10 hall at CERN (Fig. 4) The choice of the gas mixture during the
experiment was Ar + 10\% \( iC_{4}H_{10} \) (isobutane). The voltage
of the drift zone was fixed to \( HV_{drift} \) = -1000 V. The
basic elements of this test experiment were:

\begin{enumerate}
\item \textbf{Beam.} The PS accelerator at CERN delivered a secondary beam of 2 GeV/c
momentum. This beam consisted of hadrons (\( \sim  \) 60\%, mainly charged
pions) and leptons (\( \sim  \) 40\%, mainly electrons). The size of the beam
\char`\"{}spot\char`\"{} was about 10 cm in diameter and the spill duration
was 1 s with an inter-spill time of 9 s. The beam intensity stayed in the range
of \( 10^{3} \) to \( 10^{5} \) particles per spill.

\item \textbf{Beam identification.} Two Cherenkov detectors C\( _{1} \) and C\( _{2} \)
were placed upstream of the detector to identify the impinging particles. However,
only C\( _{2} \) (filled with CO\( _{2} \) at atmospheric pressure) allowed a discrimination between hadrons and leptons.

\item \textbf{Trigger.} Two plastic scintillators (Pl\( _{1} \) and Pl\( _{2}\))
with a square shape of 10\( \times \)10 cm\( ^{2} \) were inserted
along the beam line (Fig. 4) to define the main trigger of the
acquisition. Coincidences between Pl\( _{1} \) and Pl\( _{2} \)
defined a \textit{wide beam trigger}. A small plastic scintillator
Pl\( _{3} \) (1\( \times \)1 cm\( ^{2} \)) was also included in the
trigger electronics during some runs to define a \textit{narrow
beam trigger}. It was used to measure the MICROMEGAS detector
efficiency and the background induced by the passive converter.

\item \textbf{MICROMEGAS detectors.} Two MICROMEGAS detector prototypes
were placed upstream and downstream with respect to the passive
converter. Amplification of the pad signal was performed by
GASSIPLEX based electronics \cite{Sant94}. The elemental electronics card
consisted of 3 GASSIPLEX chips serving 16 channels each. These
cards (6 cards per detector) were directly attached to the detector
board. The GASSIPLEX sequential signal was digitized by the CAEN V550
VME ADC (up to 1024 channels). Operations on the GASSIPLEX and the
V550 module were synchronized via a CAEN V551A VME sequencer.

\item \textbf{Passive Converter.} During a few runs, a passive lead converter, 6 mm
thick, covering an area of 10\( \times  \)10 cm\( ^{2} \) was placed in between
the two MICROMEGAS chambers.

\item \textbf{PHOS array.} An array of 8\( \times  \)8 PHOS type crystals
was also placed at the end of the beam line but not used in our
investigation. It is mentioned here for the sake of completeness.
\end{enumerate}

\section{Detector performances}

The charge distribution collected by a single pad (Fig. 5) of the
MICROMEGAS chamber when considering the {\it wide beam trigger},
exhibits two components:

\begin{itemize}
\item For low amplitudes (50 mV) the intrinsic electronic
noise of the pad exhibits a peaked Gaussian distribution. This
corresponds to events in which beam particles fire the wide beam
trigger and hit one of the neighboring pads. The mean of
the Gaussian distribution, \( M_{G} \), results from the intrinsic noise or pedestal.
The width expressed as the standard deviation of the
distribution, \( \sigma _{G} \), results from the pad intrinsic
electronic noise, which depends on the capacitance of the pad and
on the electromagnetic environment.
\item At larger amplitudes, one observes the detector response to ionizing particles which exhibits the
usual Landau distribution reflecting the fluctuations in the number
of primary electrons created in the thin drift region. For a 3 mm
gap filled with Ar+10\% isobutane at atmospheric pressure, on
average, about 34 electrons per MIP are created. The maximum of the
Landau distribution, \( M_{L} \), reflects the gain achieved in the
amplification zone and its width the average number of primary
electrons (a smaller average number of primary electrons leading to
larger fluctuations).
\end{itemize}
The correlated noise, a noise level common to all pad signals, adds to the intrinsic noise of a pad. It can be
removed on an event-by-event basis. We defined this noise as
\begin{equation}
C_{n}=\frac{1}{N_{pad}}\sum _{|S_{i}|<3\sigma _{Gi }}^{N_{pad}}S_{i}
\end{equation}
where \( N_{pad} \) is the number of pads with a collected charge lower than
3 times the width of the noise. After removal of this correlated
noise (Fig. 5b), the charge distribution exhibits a much narrower
noise peak, offering an improved discrimination of MIP particles
from noise, and leading to an increase of the detector efficiency.

\subsection{Gain in the amplification gap}

We have studied the evolution of the Landau distribution as
a function of the micro-mesh voltage and for a constant cathode
voltage. We observe that the position of the
maximum increases with the micro-mesh voltage: an increase of 20 V
changes the position of the Landau maximum by a factor 2. The total
number of electrons collected at the anode pad is calculated as:

\begin{equation}
N_{e}^{pad}=\frac{f_{L}\times (M_{L}-M_{G})}{f_{g}\times e}
\end{equation}
where \( M_{L} \) and \( M_{G} \) are the position of the maximum
of the Landau and Gaussian distributions respectively, \( f_{g}
\) = 10 mV/fC is the gain of the GASSIPLEX (pre-amplifier and
shaper) electronics \cite{Sant94}, \( f_{L} \) is the factor to
convert the average charge from the maximum value (GEANT simulations
of the energy loss fluctuations give \( f_{L} \) = 2.5) and \( e \)
is the charge of the electron. The gain, \( G \), is calculated as:
\begin{equation}
G=\frac{N_{e}^{pad}}{N_{e}^{prim}}
\end{equation}
where \( N_{e}^{prim} \) is the number of primary electrons
generated in the conversion gap. The gain ranges from several
hundreds to several thousands for a voltage variation between 380
and 450 V (Fig. 6). The maximum achieved gain in the amplification
gap, while staying below the spark threshold, is of the order of
10\( ^{4}\).

\subsection{Detector efficiency}

The detector efficiency for MIPs was studied as a function of the
detector gain (Fig. 7 and Fig. 8).
The efficiency, \( \epsilon \), has
been therefore defined as the ratio between the integrated area of
the Landau distribution, \( L(c) \) starting from \( n\sigma _{G}\)
(where \( n\sigma _{G}\) corresponds to \( n\) times the noise width),
and of  the total integrated Landau distribution:
\begin{equation}
\epsilon =\frac{\int _{n\sigma _{G}}L(c)dc}{\int _{0}L(c)dc}
\end{equation}
where \( c \) is the induced charge in a pad. 
A direct measurement of the detector efficiency deduced from the
events registered with the \textit{small beam trigger} is in good
agreement with the value deduced using the above method. For
micromesh voltages larger than 420 V (i.e. \( G\,>\, 3.0\cdot
10^{3} \)), the measured detector efficiency is larger than 95\%.
The steep drop of the detector efficiency for voltages below 410 V
results from the fact that the gain is too low to induce a signal
larger than the nominal noise dispersion. Subtraction of the
correlated noise considerably improves the efficiency, and a 95\%
efficiency is already reached for a gain of \( 2\cdot 10^{3} \).

\subsection{Discharges in the amplification gap}

The fundamental limitation of micro-pattern gas detectors results
from the discharges induced by high particle fluxes or highly
ionizing particles. Several mechanisms have been put forward to
explain this breakdown process \cite{Font99}. In the particular
environment expected in ALICE (mainly dominated by a hadron background at low flux,
below 1 Hz/mm\( ^{2} \)), the transition to a streamer regime
followed by a electrical discharge occurs when the avalanche starts
to contain a critical quantity of a few \( 10^{7}-10^{8} \)
electrons (Raether criterion) \cite{Bres99}. Highly ionizing
particles like alpha particles loose around 500 keV of their
kinetic energy in the drift gap releasing about \(N_{e}^{prim}\) =
\( 3\cdot 10^{4} \) primary electrons, whereas MIPs loose only around
1 keV (generating around 30 primary electrons). Alpha particles
will, therefore, induce a discharge in the MICROMEGAS for gains
above several \( 10^{3}\) (leading to \( N_{pad} \) larger
than \( 10^{7} \)). The same effect is observed when the MICROMEGAS
detector is irradiated with hadron MIPs, like charged pions. In
this case, nuclear interactions with nuclei of the gas mixture in
the drift gap could be at the origin of the induced discharges,
since a strong dependence with the average atomic number of the gas
mixture has been observed \cite{Micr00,Rebo99}. Although MICROMEGAS
detectors are very resistant to sparks, a discharge will induce a
non-negligible dead time over the whole detector active area,
leading to a reduction of the effective efficiency.

We have measured (Fig. 9) the discharge probability per ionizing
particle as a function of the  detector gain. We obtain discharge
probabilities roughly between \( 10^{-6} \) and \( 10^{-5} \) for
different prototypes. The discharge probability does not depend on
the beam intensity, as it could be expected. However, since the
beam intensity was measured by Pl\( _{1} \) and Pl\( _{2} \)
coincidences, impinging-particle trajectories outside of the
plastic active area and passing through the MICROMEGAS detector,
were not counted. This induces an asymmetric systematic error which
would tend to slightly decrease the measured discharge probability.

Central heavy ion collisions at LHC will induce a flow of charged
particles of the order of 50 charged particles per m\( ^{2}\) in
PHOS \cite{PHOS99}. Taking a conservative value of 8000 central
collisions per second, the charged particle rate will be of the
order of 50 KHz per chamber. If the detector operates at gains of
2000 (leading to efficiencies larger than 95\%), we expect a spark
probability of \( 3\cdot 10^{-6} \) (Fig. 9), i.e. a spark rate of
0.15 spark/chamber/second. Taking into account the fact that the
dead time induced by a spark is less than 10 ms
\cite{Micr00,Rebo99}, the dead-time induced by sparks will be as low as 0.15\%.
In this respect, the particle induced discharges are not a
handicap of the MICROMEGAS detectors, when the chambers
operate at modest gains.

In addition, recent measurements
\cite{Micr00,Rebo99}, exhibit a strong dependence of the spark
probability per ionizing particle on the average atomic number of
the gas mixture, following a \( Z^{-4}\) dependence. Therefore,
selecting a gas mixture with Ne gas should lead to an additional
reduction of the spark probability by more than one order of
magnitude.

\subsection{Background induced by the passive converter}

The background induced by the lead passive converter in the
MICROMEGAS pad chambers has also been studied. Using the
\textit{narrow beam trigger} (see Fig. 10 where almost no noise
peak is apparent) we have measured the charge distribution in the
neighboring pads without and with the lead passive converter. We
observe the following (Fig. 11):

\begin{itemize}
\item without converter, 2\% of MIPs induce a signal larger than
the nominal noise in a neighboring pad;
\item with converter, only 6\% of MIP particles induce a signal
in a neighboring pad.
\end{itemize}
We conclude that the passive converter induced background is
relatively small, and the detector occupancy of the MICROMEGAS pad
chambers is hardly increased.

\subsection{Pre-shower response function to electrons and hadrons}

In this last section, we present the response function of the
pre-shower detector to hadrons and electrons of 2 GeV/c momentum.
The discrimination between the hadron and electron beam was done by
means of the Cherenkov counter C\( _{2} \) (Fig. 4). Comparisons of
the experimental response functions with the results of GEANT3.21
simulations of the pre-shower detector were also performed \cite{Aliroot99}.

\begin{itemize}
\item \textbf{Hadron beam} (Fig. 12): The PC detector exhibits almost the
same Landau-like distribution with and without the converter.
Fluctuations of the deposited energy in the thin conversion gas gap
were calculated using the PAI model implemented in GEANT. This
simulation reproduces relatively well the shape of the Landau-like
distribution, although GEANT underestimates the tail of the
measured distribution.
\item \textbf{Electron beam} (Fig. 13): Without converter, the PC
detector exhibits the same Landau-like distribution observed with
the hadron beam. However, by including the converter (1.07 X\( _{0}
\)) in the setup, a wider distribution is observed: on average 65\%
of the impinging electrons develop an electromagnetic shower inside the converter.
The
secondary particles produced in the shower will result, on average,
in a higher energy deposit in the gas cell associated to the
anode-pad leading to a broader charge distribution. GEANT3.21
simulations of the deposited energy by 2 GeV/c electrons describe
rather well the observed distribution.
\end{itemize}
We conclude that the hadron response function is not significantly modified by
the passive converter. However, electrons exhibit a very different
response function in the presence of the passive converter, due to
the initial development of electromagnetic showers. The deposited
energy distribution in PC detector is wider, and this observation
could be exploited to improve the electron and positron identification power of the
PHOS detector.

\section{Conclusions and perspectives}

We have developed a large-area MICROMEGAS detector with anode pads
read-out. Prototypes have been irradiated with a 2 GeV/c momentum
beam of electrons and hadrons. We have
studied the pad response function, the gain in the amplification
gap and the detector efficiency as a function of the electric
field. The measured performances are in good agreement with those
obtained in other MICROMEGAS developments. We have investigated the
discharge probability per incident MIP particle, concluding that
the induced discharges do not represent a real handicap to operate
MICROMEGAS chambers in the environment expected in central
heavy-ion collisions at LHC energies. A pre-shower detector design
based on a sandwich of two MICROMEGAS chambers with a passive Pb
converter in between, has been proven to be a promising option for
the improvement of PHOS detection capabilities. The measured
background induced by the converter in the MICROMEGAS chambers is
very small. The pre-shower response function of hadrons and
electrons presents various facets worth exploiting further. We
observe that the hadron beam induces, in both gas detectors, the
typical Landau-like distribution of the collected charge. The
electron beam not only induces the typical Landau-like distribution
in the CPV detector of the pre-shower but also exhibits a much
wider charge distribution in the PC detector, signing the initial
development of the electromagnetic shower.

Following the results of the tests discussed in the present
article, an improved version of a PPSD has been designed.
The new detector  has a larger active area (380\(\times \)380 mm\( ^{2} \)),
a thicker conversion gap (6 mm) better suited for the use of a gas mixture like Ne with 5-10\% of
CO\(_{2} \) and a 200 $\mu$m amplification gap \cite{Mart00}.
These improvements (larger amplification gap and lighter gas mixture),
will lead to allow for a reduction of the spark probability
by a factor 20 to 50, and therefore lead to a spark rate of 10 sparks/chamber/hour for the ALICE hadronic environment.
In addition, the size of the pad has been reduced to 11\( \times\)11 mm\( ^{2} \)
in order to reach the spatial resolution needed to determine the shower vertex.

\section{Acknowledgments}

We thank M.~Ippolitov and PHOS collaborators for their help during the beam
test at PS (CERN).
The technology for manufacturing the circuit board has
been efficiently developed by A. Gandi and R. de Oliveira
(CERN-EST-MT).
We would like to acknowledge here the fruitful
discussions with Y.~Giomataris and Ph.~Rebourgeard at DAPNIA
(Saclay), and  with P. Lautridou, L. Luquin and M. Labalme of
Subatech (Nantes) during the development of the MICROMEGAS prototypes.
This work was supported in part by the ``Conseil R\'egional de la
R\'egion des Pays de la Loire'', France.

\newpage

\begin{figure}
{\par\centering \resizebox*{0.9\textwidth}{!}{\rotatebox{270}{\includegraphics{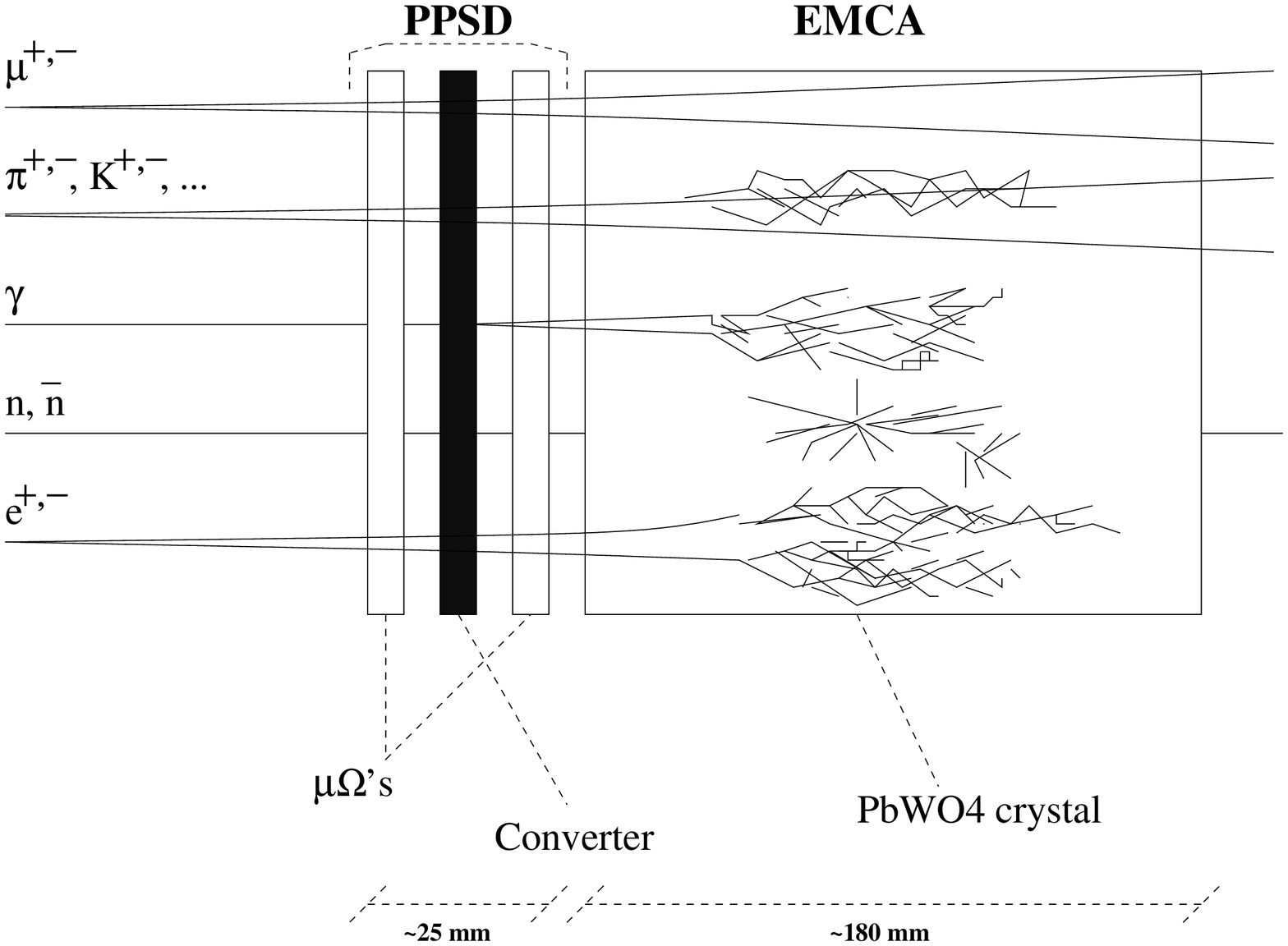}}} \par}
\caption{Schematic view of the Pre-shower detector (PPSD) in front of the
PHOS electromagnetic calorimeter (EMCA). \( \mu\Omega \) stands for
MICROMEGAS gas counters, CPV for the charged particle veto chamber and
PC for the photon conversion chamber.}
\end{figure}

\newpage
\begin{figure}
{\par\centering \resizebox*{0.85\textwidth}{!}{\rotatebox{0}{\includegraphics{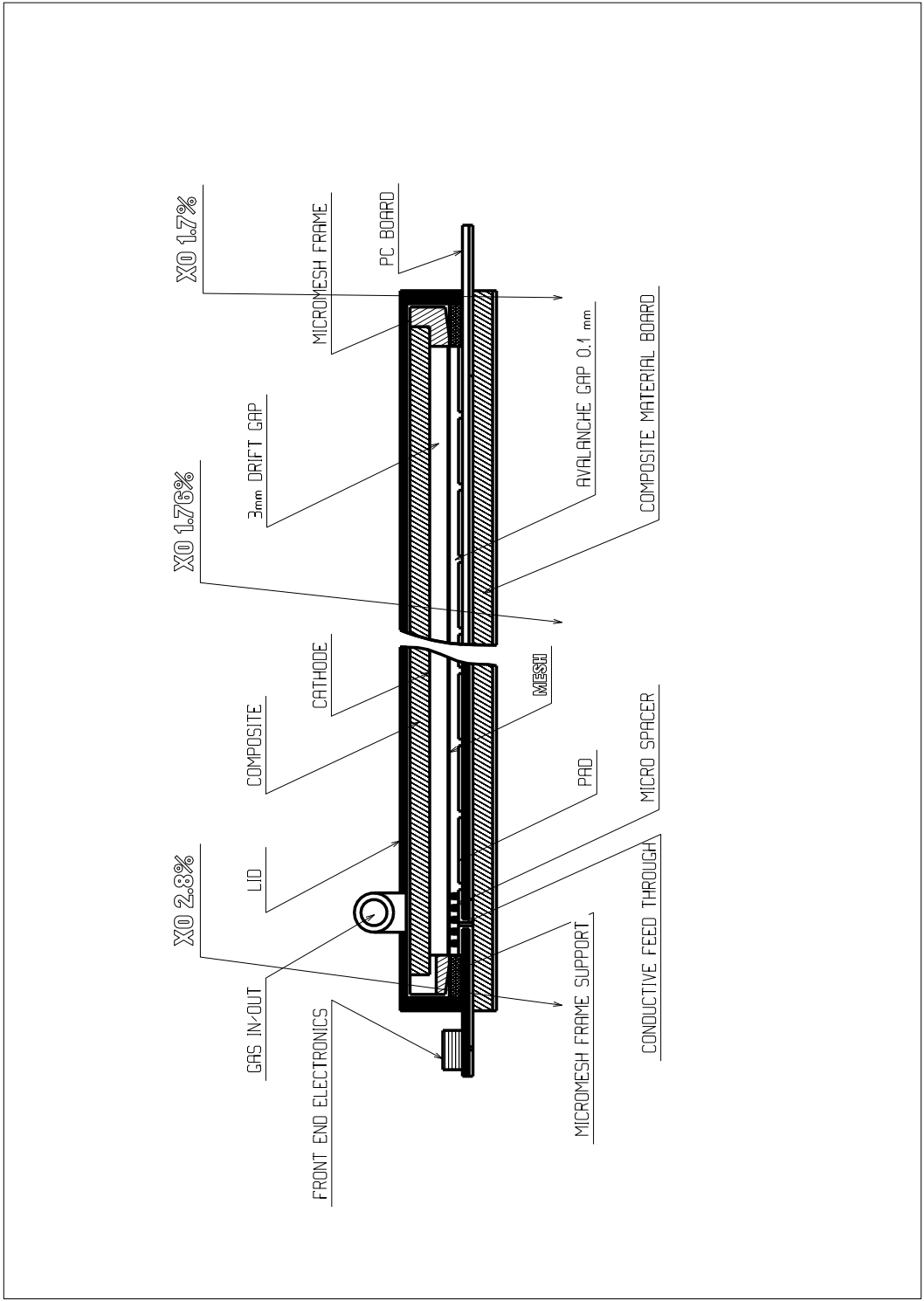}}} \par}
\vspace{1.5cm}
\caption{Principle design of our MICROMEGAS detector prototype: side view of an assembled
detector.}
\end{figure}

\newpage
\vspace{2.5cm}
\begin{figure}
{\par\centering \resizebox*{0.75\textwidth}{!}{\includegraphics{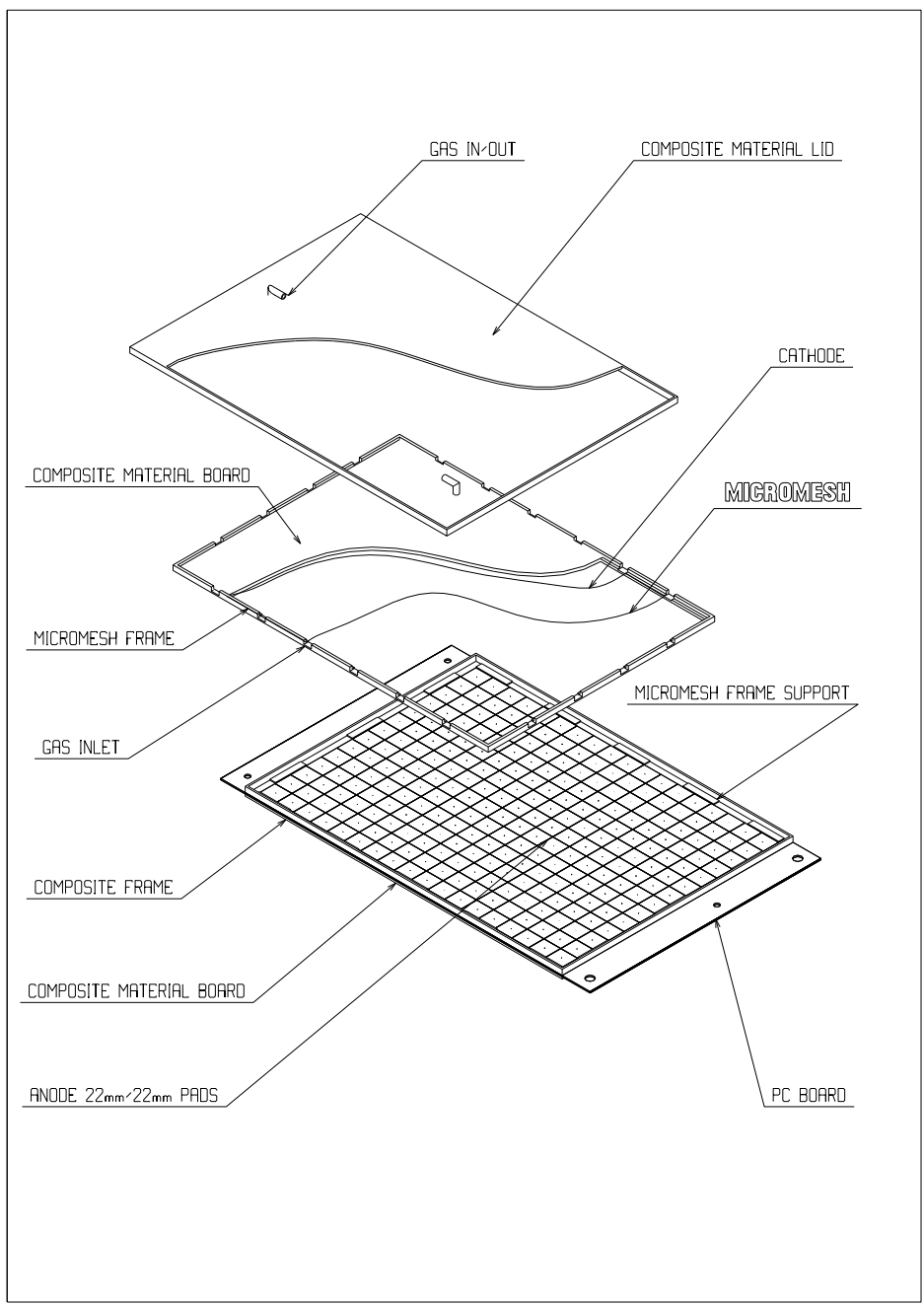}} \par}
\vspace{4.cm}
\caption{Principle design of our MICROMEGAS detector prototype: enlarged view showing
the various components.}
\end{figure}

\newpage
\vspace{2.5cm}
\begin{figure}
{\par\centering \resizebox*{0.6\textwidth}{!}{\rotatebox{270}{\includegraphics{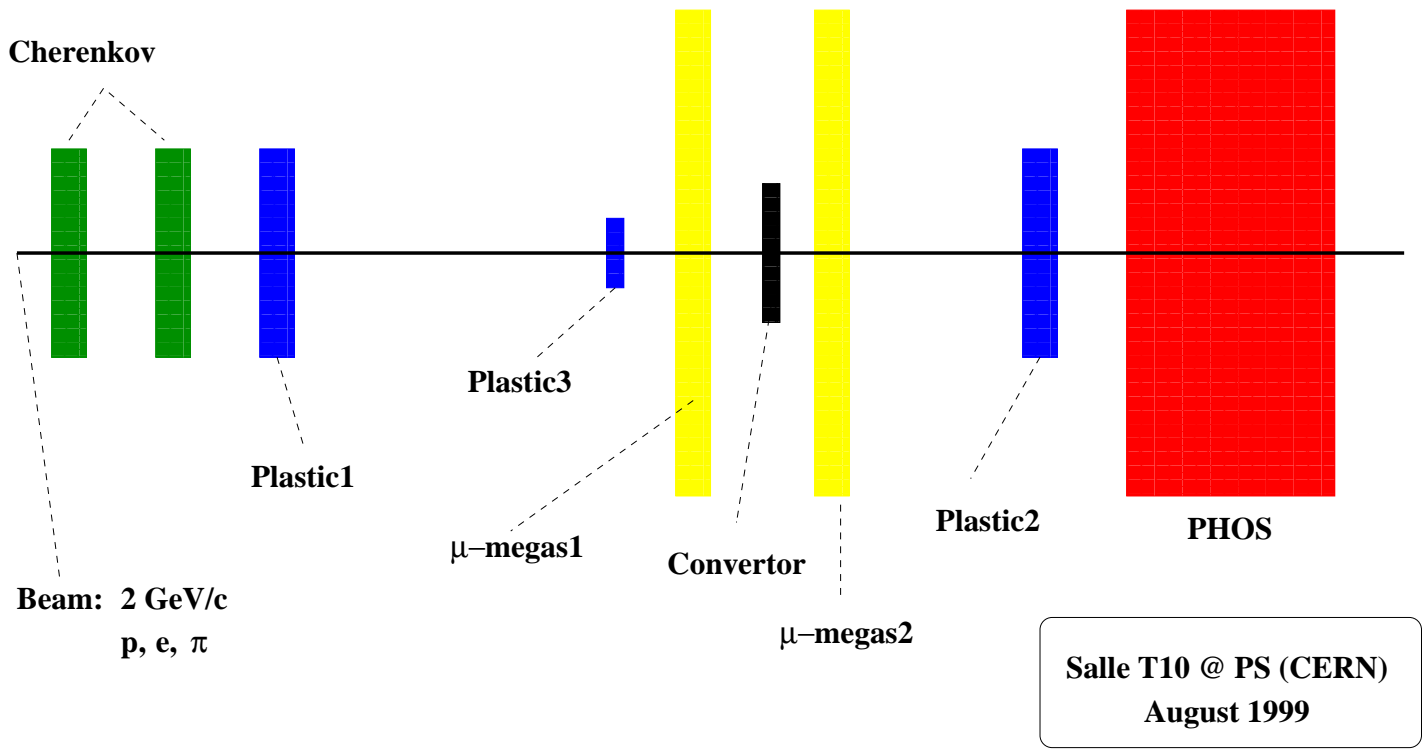}}} \par}
\vspace{2.5cm}
\caption{Experimental set-up during the in-beam tests.}
\end{figure}

\newpage
\vspace{2.5cm}
\begin{figure}
{\par\centering \resizebox*{1\textwidth}{!}{\includegraphics{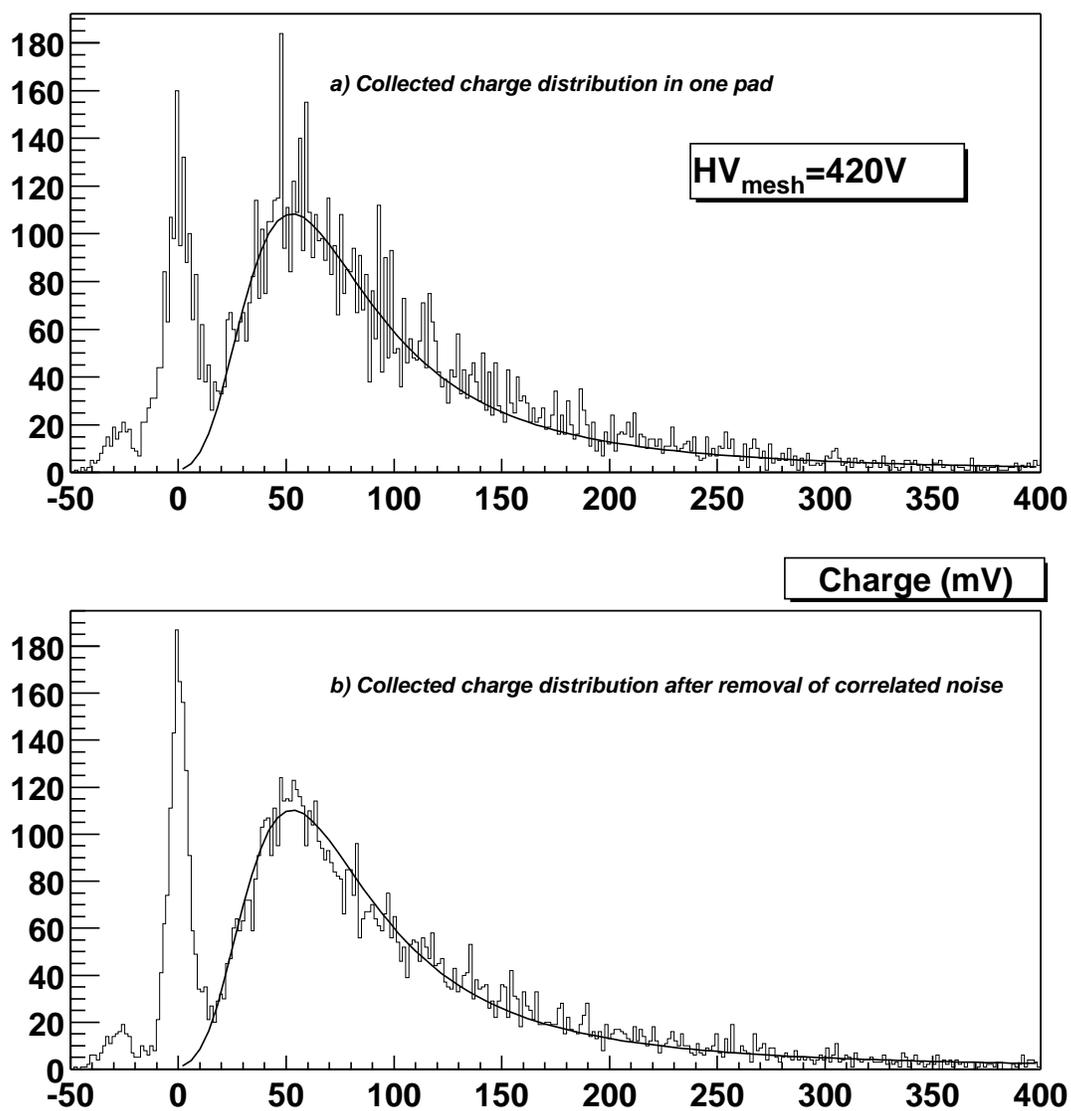}} \par}
\caption{a) Charge distribution collected on a single pad induced by particles
(electrons and hadrons) of 2 GeV/c momentum impinging on a
MICROMEGAS chamber. The solid line is a fit of the signal to a
Landau distribution. b) The same charge distribution after removal
of the correlated noise.}
\end{figure}

\newpage
\vspace{2.5cm}
\begin{figure}
{\par\centering \resizebox*{1\textwidth}{!}{\includegraphics{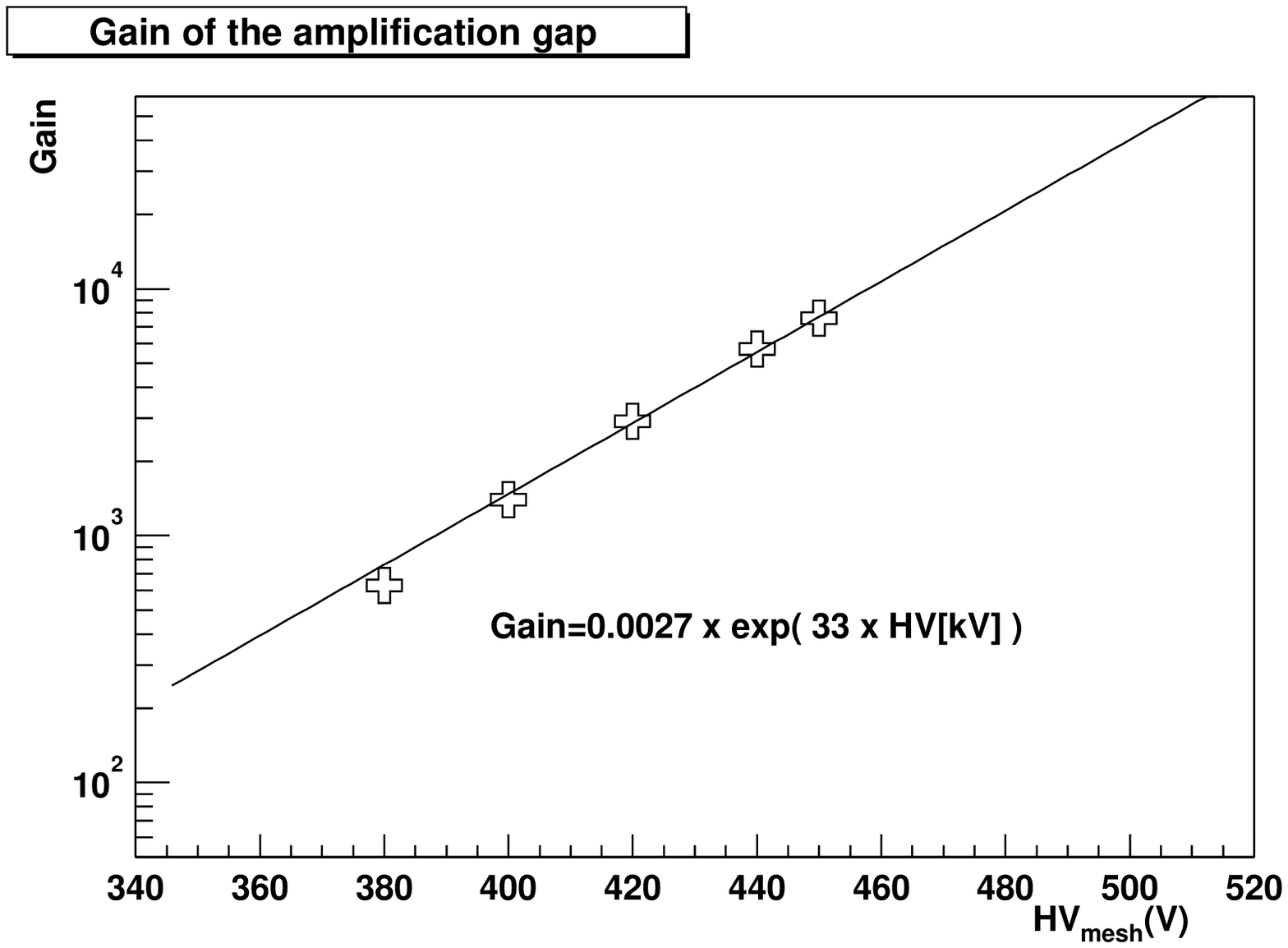}} \par}
\caption{Gain in the amplification gap as a function of the voltage applied on the micro-mesh. The solid line is to guide the eyes.}
\end{figure}

\newpage
\vspace{2.5cm}
\begin{figure}
{\par\centering \resizebox*{1\textwidth}{!}{\includegraphics{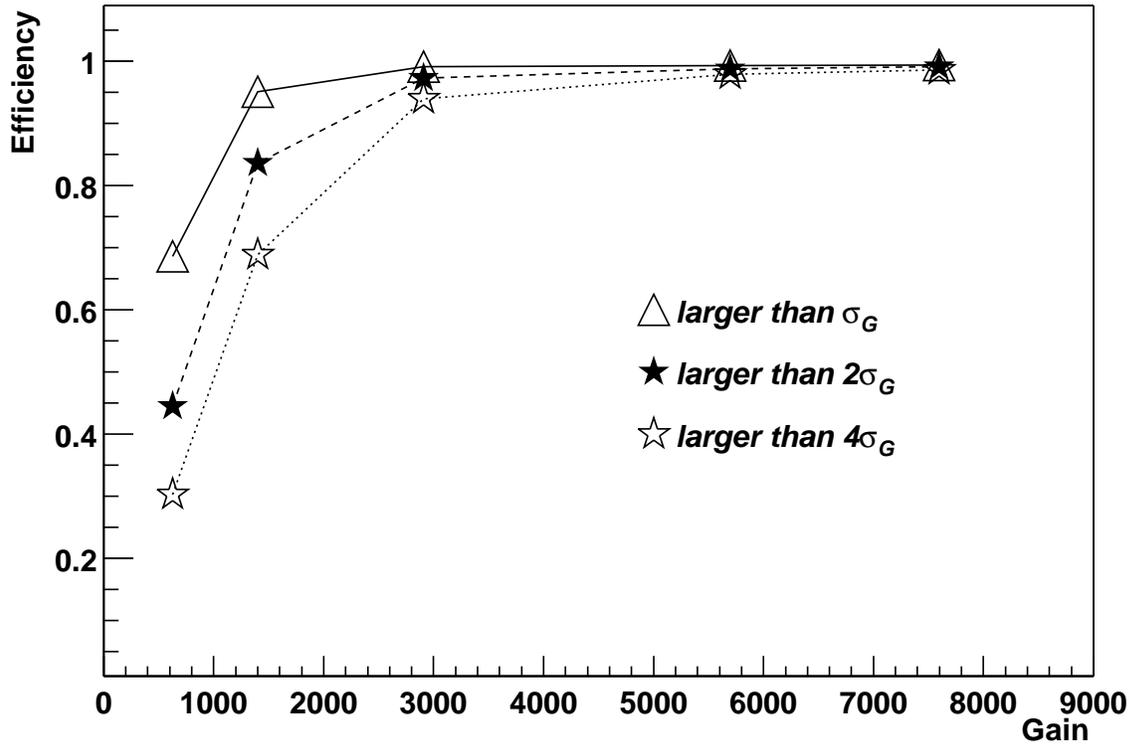}} \par}
\caption{Detector efficiency for MIPs as a function of the amplification gain for different
values of the noise gaussian width \(\sigma _{G}\). The lines drawn through the symbols are to guide the eyes.}
\end{figure}

\newpage
\vspace{2.5cm}
\begin{figure}
{\par\centering \resizebox*{1\textwidth}{!}{\includegraphics{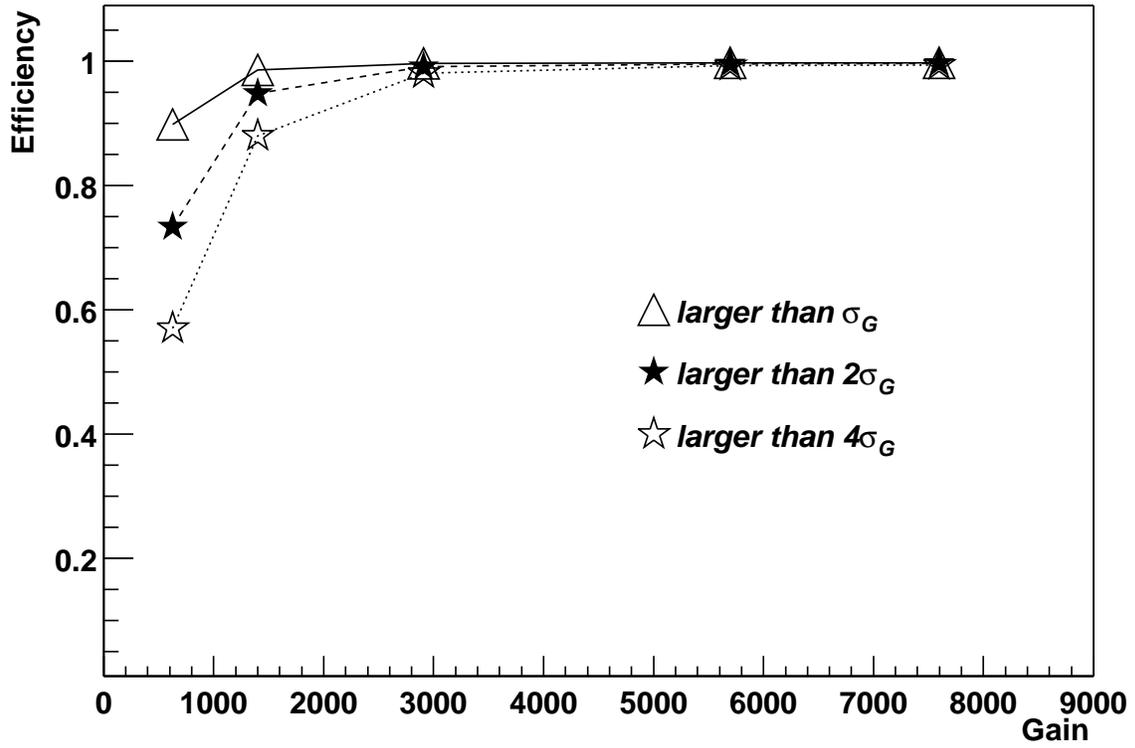}} \par}
\caption{Detector efficiency for MIP as a function of the amplification gain, after
removal of the correlated noise \( C_{n} \).  The lines drawn through the symbols are to guide the eyes.}
\end{figure}

\newpage
\vspace{2.5cm}
\begin{figure}
{\par\centering \resizebox*{1\textwidth}{!}{\includegraphics{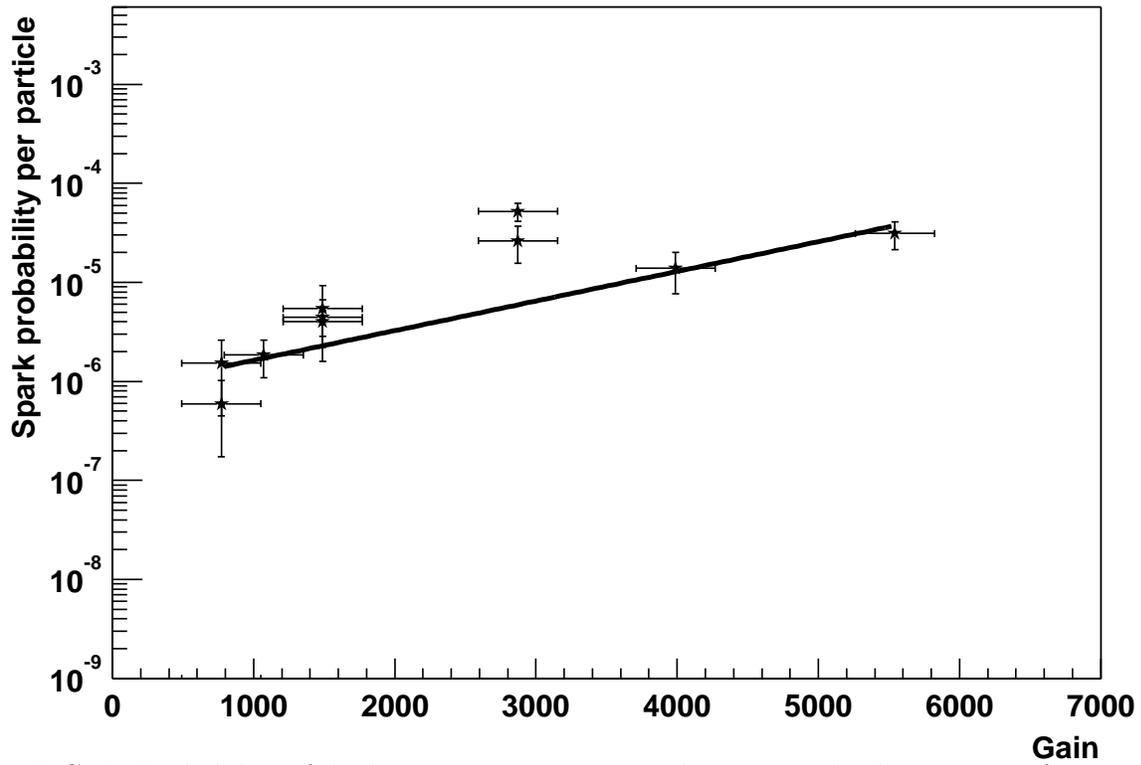}} \par}
\caption{Probability of discharge per ionizing particle crossing the detector as a function of the gain.
The lines drawn through the symbols are to guide the eyes.}
\end{figure}

\newpage
\vspace{2.5cm}
\begin{figure}
{\par\centering \resizebox*{1\textwidth}{!}{\includegraphics{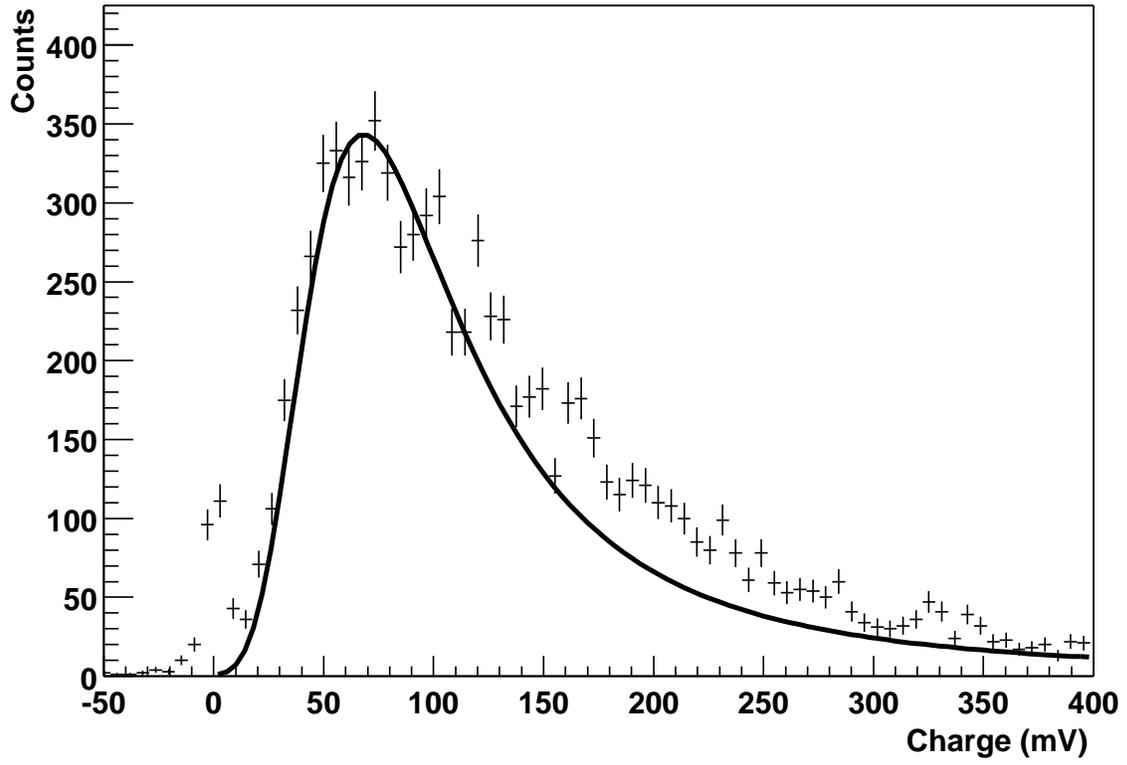}} \par}
\caption{Pad response function for MIPs in the case of the narrow beam trigger.}
\end{figure}

\newpage
\vspace{2.5cm}
\begin{figure}
{\par\centering \resizebox*{1\textwidth}{!}{\includegraphics{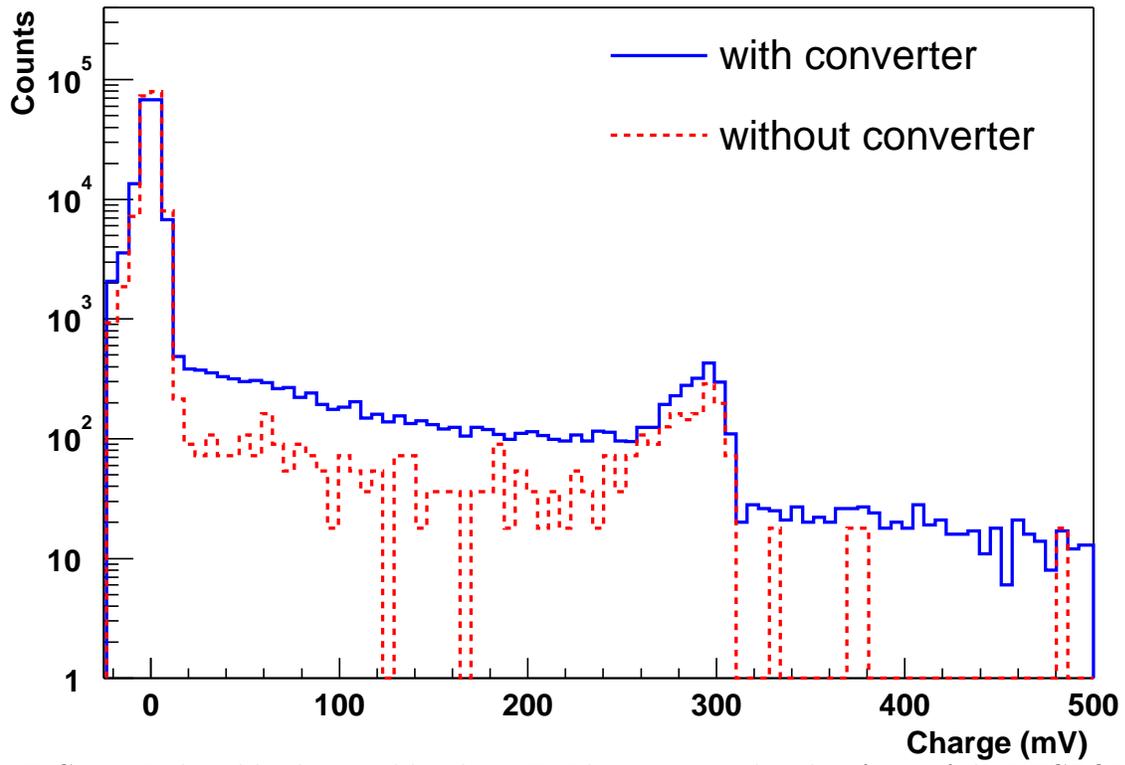}} \par}
\caption{Induced background by the 1 \( X_{0} \) Pb converter placed in front of the MICROMEGAS chamber.}
\end{figure}

\newpage
\vspace{2.5cm}
\begin{figure}
{\par\centering \resizebox*{1\textwidth}{!}{\includegraphics{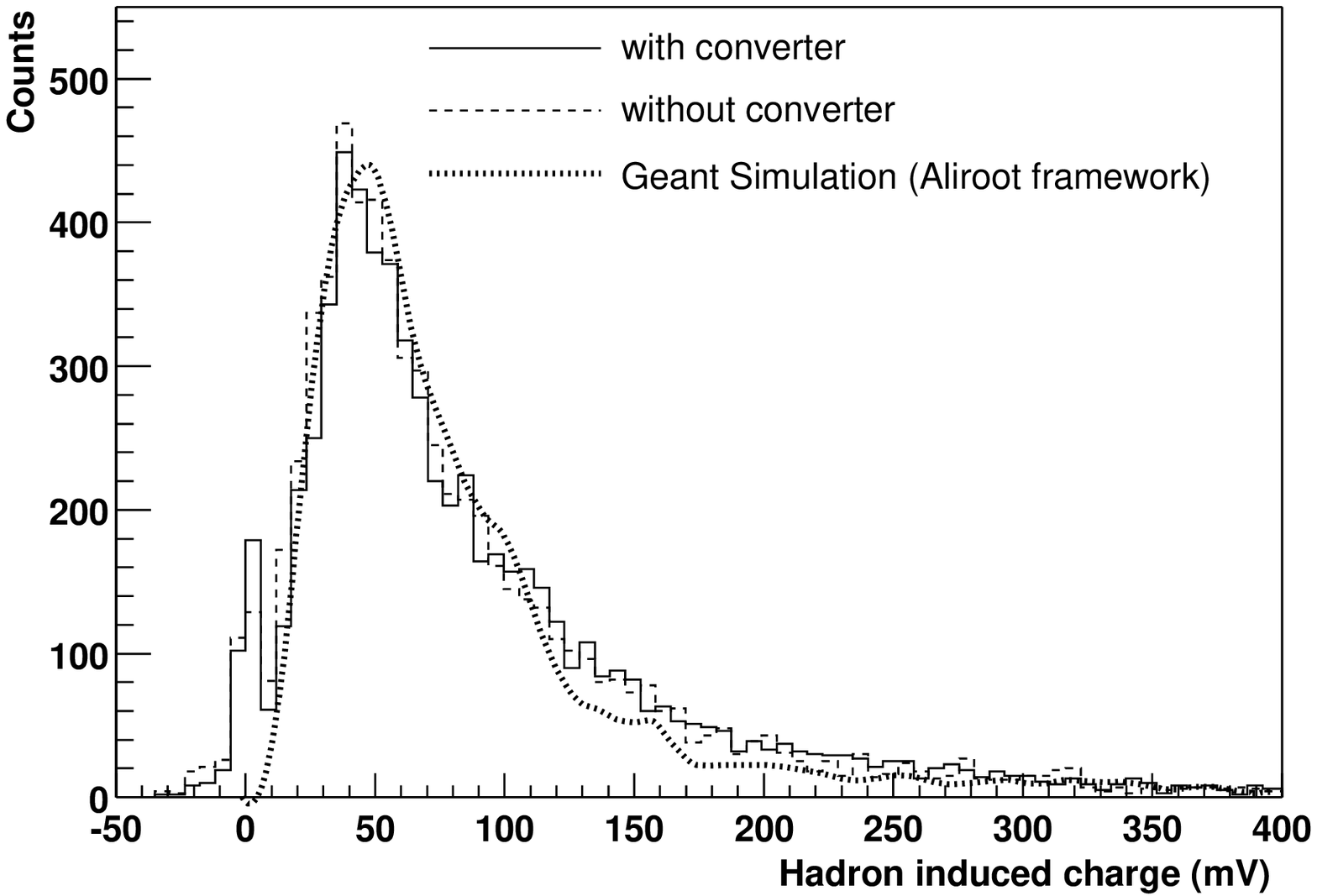}} \par}
\caption{Measured pad response function for hadrons with and without converter, and a GEANT simulation.}
\end{figure}

\newpage
\vspace{2.5cm}
\begin{figure}
{\par\centering \resizebox*{1\textwidth}{!}{\includegraphics{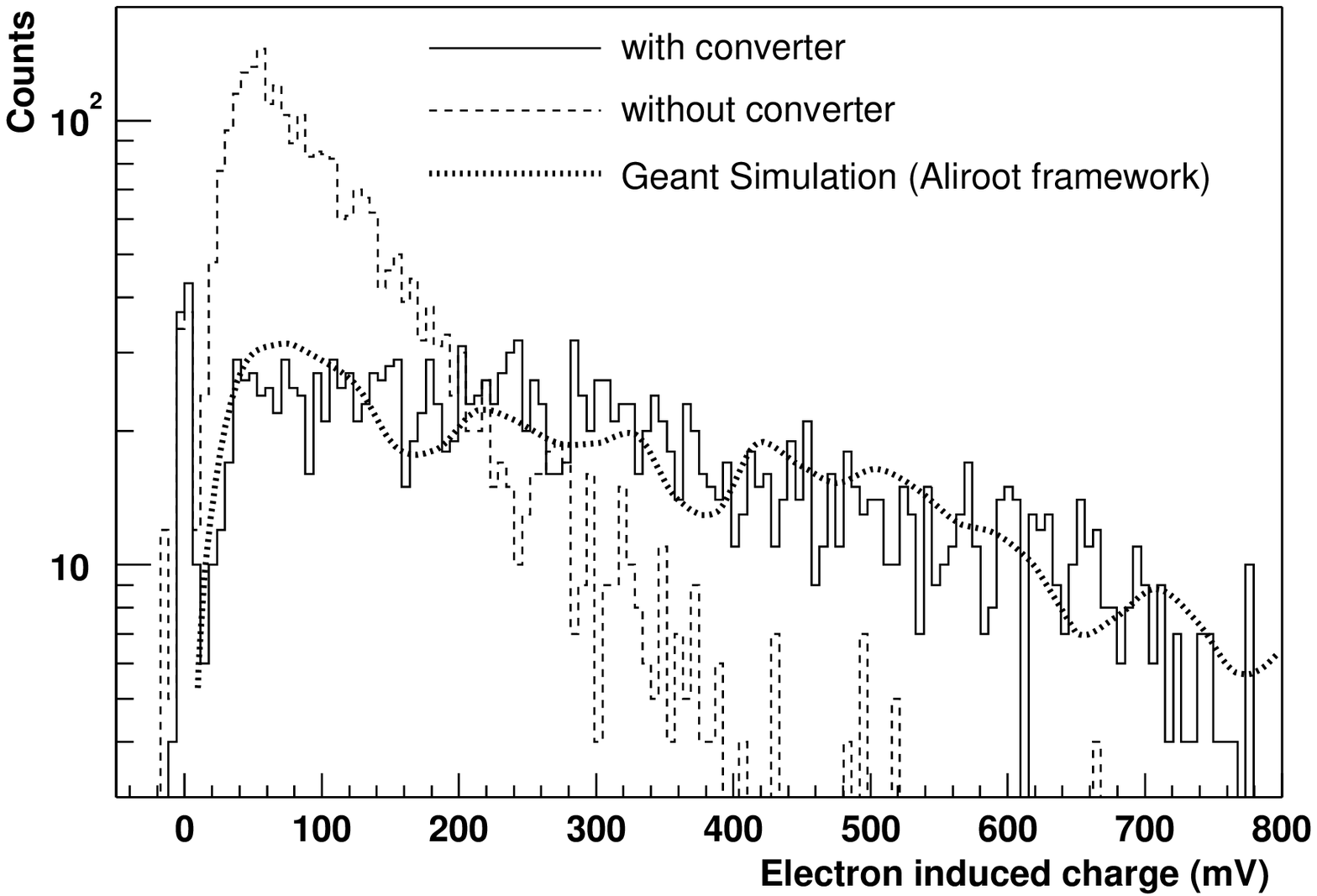}} \par}
\caption{Measured pad response function for electrons with and without converter, and a GEANT simulation.}
\end{figure}

\end{document}